\documentclass[12pt]{iopart}

\usepackage{graphics}
\usepackage{epsfig}
\usepackage{iopams}

\usepackage{color}
\usepackage{pdfsync}

\usepackage{ulem}

\begin{document}

\title{Field dependence of the Eu$^{2+}$ spin relaxation in EuFe$_{2-x}$Co$_{x}$As$_{2}$ }

\author{F.A. Garcia, A. Leithe-Jasper, W. Schnelle, M. Nicklas, H. Rosner,
and J. Sichelschmidt}

\address{Max Planck Institute for Chemical Physics of Solids, 01187 Dresden, Germany}

\date{\today}

\begin{abstract}
The layered compound EuFe$_{2}$As$_{2}$ is an interesting model system to investigate the effects of well defined local Eu$^{2+}$ 4$f$ states on the itinerant electronic and magnetic properties of the FeAs layers. To address this subject, we investigated the series EuFe$_{2-x}$Co$_{x}$As$_{2}$ ($0.1\leq x\leq0.75$) by electron spin resonance (ESR) of Eu$^{2+}$ to probe the spin dynamics of the itinerant subsystem. We relate the results to dc-susceptibility measurements and band structure calculations. As a consequence of the weak coupling between the local and itinerant subsystems we found that the spin relaxation is well understood in terms of the exchange coupling among the local Eu$^{2+}$ spins. A pronounced field dependence of the Eu$^{2+}$ spin relaxation demonstrates the direct influence of magnetic fluctuations at the Fe$_{2-x}$Co$_{x}$As$_2$ layers.
\end{abstract}

\maketitle

\section{Introduction}

The discovery of superconductivity in iron-based pnictides \cite{kamihara_iron-based_2008,zhi-an_superconductivity_2008}
sparked a tremendous surge of interest and research activities. This
class of materials displays a high structural and chemical flexibility
comprising several closely related structure types based on common
layered Fe-pnictogen building blocks. Depending on charge-carrier
doping, external or chemical pressure they all show a subtle interplay
among structural transitions, antiferromagnetism (AFM) and superconductivity
\cite{johnston_puzzle_2010}. The relevance of magnetic fluctuations
for the stabilization of the superconducting state has been put forward
for unconventional superconductors which are found among a wide range
of compound families \cite{monthoux_superconductivity_2007,saxena_superconductivity_2000,le_tacon_intense_2011}.
Their role in superconducting iron-pnictides are now subjected to
intense scrutiny. In this context, the compound EuFe$_{2}$As$_{2}$
(with the body centered tetragonal ThCr$_{2}$Si$_{2}$- type of structure,
where layers of Fe$_{2}$As$_{2}$ alternate with Eu layers along
the c-axis) was presented as an important model system to study the
interplay between well defined 4$f$ magnetic states and the electronic
properties of the iron-arsenide layer \cite{jeevan_electrical_2008,ren_antiferromagnetic_2008,kasinathan_afe2as2_2009}.

EuFe$_{2}$As$_{2}$ is an intermetallic compound which undergoes
both a spin density wave (SDW) type antiferromagnetic phase transition at $T_{\rm N}^{\mathrm{SDW}}=195$
K, associated with the itinerant states of the FeAs subsystem, and
a localized AFM ordering taking place at $T_{\rm N}^{\mathrm{Eu}}=22$
K, associated with the Eu 4$f$ local moments subsystem \cite{jeevan_electrical_2008,ren_antiferromagnetic_2008}.
The magnetic structure of the local moment subsystem consists of Eu$^{2+}$
layers in a ferromagnetic order, stacked antiferromagnetically along
the $c$-axis \cite{xiao_magnetic_2009}. It has been shown that
upon substitution on any of the Eu (K substitution) \cite{jeevan_high-temperature_2008},
Fe (Co substitution) \cite{ying_electron_2010} and As (P substitution)
sites \cite{jeevan_interplay_2011,ren_superconductivity_2009}, the
itinerant AFM state is suppressed and the onset of a superconducting
phase is observed. Similar behavior is also found when the host compound
is subjected to external pressure, that is also claimed to lead to
the appearance of reentrant superconductivity \cite{miclea_evidence_2009,kurita_phase_2011}.

Notwithstanding the large effects observed in the itinerant AFM phase,
the energy scale of the local moment interaction, as compared to the
energy scale of the itinerant subsystem, has been observed to be only
slightly modified in all of these substitutional studies. This suggests
a weak coupling between the local 4$f$ electronic states and the
electronic and magnetic properties of the FeAs layers. Further indication
in this direction is also drawn from the study of the rather weak
suppression of the superconducting state by the inclusion of Eu in
optimally doped Sr$_{1-y}$Eu$_{y}$Fe$_{2-x}$Co$_{x}$As$_{2}$
\cite{hu_phase_2011}, from the lack of signatures on the electronic
structure of EuFe$_{2}$As$_{2}$ when the Eu$^{2+}$ subsystem undergoes
the AFM transition \cite{zhou_high-resolution_2010} and from magnetotransport
studies \cite{terashima_magnetotransport_2010}. 

The nature of the local moment ordering, however, is changed from
AFM to ferromagnetic (FM) upon substitution of As by P \cite{jeevan_interplay_2011}
and possibly also for Fe by Co \cite{nicklas_competition_2011}.
In the case of P substitution, it seems clear that the change from
AFM to FM order is related to both structural effects and substitutional
effects \cite{jeevan_interplay_2011}. This is in contrast to the
Co substitution, where the change in the electron filling is supposed
to be the dominant effect, allowing a more controlled investigation. 

In searching a comprehensive description of the magnetic and electronic properties of the FeAs
layers, the electron spin resonance of Eu$^{2+}$ has proven to be a suitable tool to provide information on the spin dynamics in EuFe$_{2}$As$_{2}$ \cite{dengler_strong_2010}. In this respect, a proper understanding of the Eu$^{2+}$ spin relaxation is essential, and especially needs to be considered for \textit{concentrated} Eu$^{2+}$ ions like in EuFe$_{2-x}$Co$_{x}$As$_{2}$. In case of Eu$^{2+}$ ions \textit{diluted} in a metallic environment, the spin relaxation is determined by the exchange scattering between the local and itinerant spins (the Korringa process) and should be sensitive to the spin dynamics of both \cite{barnes_theory_1981}. However, in concentrated magnetic compounds, the electronic spin relaxation is expected to
be determined more by the exchange coupling among the local spins
\cite{van_vleck_dipolar_1948,huber_electron_1976} which is sensitive to an externally applied magnetic field. Therefore, a systematic investigation of the field dependence of the ESR in EuFe$_{2-x}$Co$_{x}$As$_{2}$ and an interpretation in terms of concentrated Eu$^{2+}$ ions provide a unique tool to evaluate the relation between the local moment magnetism and the magnetic and electronic properties of the itinerant subsystem.

To this aim, along with dc-susceptibility measurements, we present
the Eu$^{2+}$ ESR at three different magnetic fields, corresponding
to resonance frequencies of $1.1$ GHz ($L$-band), $9.4$ GHz ($X$-band)
and $34$ GHz ($Q$-band). We shall discuss a significant reinterpretation
of the available ESR data \cite{dengler_strong_2010,pascher_magnetic_2010,ying_electron_2010,garcia_doping_2011},
that do not include an investigation on the field dependence of the
relaxation. The field dependence of the relaxation is a strong piece
of evidence that the Eu 4$f$ local moments are coupled with the Fe$_{2-x}$Co$_{x}$As$_2$
itinerant magnetic fluctuations. Furthermore, we have also perfomed
band structure calculations to guide our discussion on the nature
of the Eu$^{2+}$ spin relaxation.

The local moment relaxation is sensitive to the exchange coupling between the local spins by virtue of the so-called bottleneck effect \cite{barnes_theory_1981}. It occurs when the conduction electron-lattice scattering rate ($1/T_{\mathrm{ceL}}$) is  smaller than, or comparable to, the conduction electron-magnetic ion ($1/T_{\mathrm{ceI}}$) scattering rate. In this situation,
after being scattered by the magnetic center (the Korringa process),
the conduction electrons, instead of dissipating energy to the lattice,
give this energy back to the magnetic ion system (Overhauser process).
As a consequence, one may probe a suppressed Korringa process \cite{barnes_theory_1981,rettori_dynamic_1974,rettori_electron-spin_1973},
since the slow $1/T_{\mathrm{ceL}}$ will modulate the Korringa
rate. However, it may also occur that the relaxation simply does not
reflect the exchange scattering between the local moments and the
conduction electrons, being determined primarily by the exchange coupling
between the Eu$^{2+}$ ions \cite{van_vleck_dipolar_1948,huber_electron_1976}. 

Even in the latter scenario the properties of the Fe$_{2-x}$Co$_{x}$As$_2$
itinerant subsystem can be deduced from the local moment relaxation,
since the coupling among local spins is realized through the Ruderman-Kittel-Kasuya-Yosida
(RKKY) indirect exchange interaction. Hence, while mediating the exchange
coupling of the local subsystem, the itinerant subsystem manifest
its properties in the local moment relaxation. This interplay qualifies
the Eu$^{2+}$ ESR as a suitable probe for the spin dynamics taking
place in the Fe$_{2-x}$Co$_{x}$As$_2$ layers.

\section{Methods}

Details of the synthesis of polycrystalline samples of EuFe$_{2-x}$Co$_{x}$As$_{2}$
($0.1\leq x\leq0.75$) are given in Ref. \cite{nicklas_competition_2011}.
The actual content of Co was determined by energy dispersive X-ray scattering
(EDX) and are very close to the nominal values. The physical properties
of the samples used in this experiment are presented in Ref. \cite{nicklas_competition_2011}.
The ESR measurements were carried out using a Bruker Elexsys 500 spectrometer
for $L$-band ($1.1$ GHz), $X$-band ($9.4$ GHz) and $Q$-band ($34$
GHz) frequencies in the temperature intervals $30\leq T\leq120$ K
for the $L$-band and $4.2\leq T\leq300$ K for both $X$-band and
$Q$-band measurements. 

The samples used in the ESR measurements were sieved in a fine powder
to achieve grain size homogeneity. In the whole temperature interval,
the spectrum consisted of an asymmetric exchange narrowed single broad
line. The ESR parameters were obtained through the fitting of the
observed spectra according the formula given in Refs. \cite{joshi_analysis_2004, wykhoff_local_2007} which takes also the effect of a counter resonance in broad ESR lines into account.
The parameters included in the fitting are the ESR linewidth $\Delta H$,
the resonance field $H_{\mathrm{res}}$, the spectrum amplitude and
the lineshape parameter $\alpha=D/A$, which takes into account the
dispersion to absorption ratio of the microwave radiation when it
probes a metallic surface ($D=0$ in insulators). In the whole temperature
interval, the best fitting was always given by this procedure, although
we have also tried to search for contributions given by unresolved
crystal field (CF) effects and other magnetic anisotropy effects which
could be present in the powder spectra. The dc-susceptibility of the
samples was measured using a commercial MPMS SQUID dc magnetometer
(Quantum Design).

Scalar-relativistic density functional (DFT) electronic structure
calculations were performed using the full-potential FPLO code \cite{opahle_full-potential_1999},
version fplo9.01-35. For the exchange-correlation potential, within
the local density approximation (LDA) the parametrization of Perdew-Wang \cite{perdew_accurate_1992}
was chosen. To obtain precise band structure information, the calculations
were carried out on a well converged mesh of 4096 $k$-points (16x16x16
mesh, 405 points in the irreducible wedge of the Brillouin zone).
The partial Fe substitution with Co was modeled within the virtual
crystal approximation (VCA) as implemented according to Ref. \cite{kasinathan_electronic_2012}.
Since the structural changes upon partial Co substitution are small
in the investigated concentration range, the experimental structural
data for the undoped tetragonal EuFe$_{2}$As$_{2}$ compound (space
group I4/mmm, a $=3.916 \AA$, c $=12.052 \AA$ and z$_{\mathrm{As}}$$=0.3625$) \cite{uhoya_anomalous_2010}
were used throughout the calculations. The strong Coulomb repulsion
for the Eu 4$f$ states were treated within the LSDA+$U$ approximation
with $U=8$ eV as a typical value for Eu. The resulting density of
states (DOS) is essentially unchanged for a variation of $U$ within
the physically relevant range.

\section{Results and Discussion}

\begin{figure}
\begin{centering}
\includegraphics[scale=0.6]{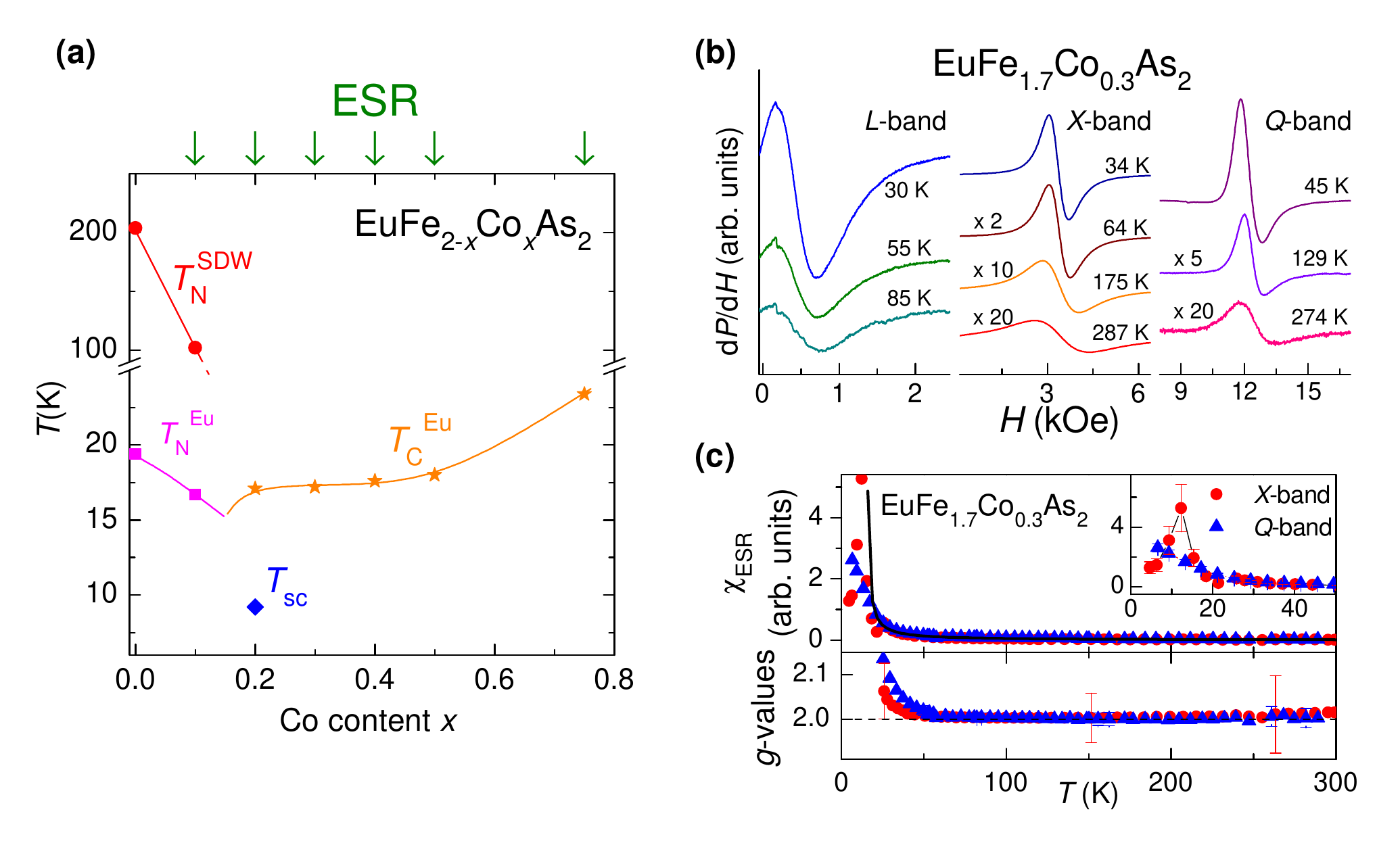}
\par\end{centering}

\caption{(Color online) General characteristics of the investigated samples. (a) Temperature ($T$ ) vs  $x$ phase diagram for the EuFe$_{2-x}$Co$_{x}$As$_{2}$ series. Arrows indicate the samples measured by ESR. (b) Survey of representative ESR spectra for $x=0.3$ taken at  $L$-band ($1.1$ GHz), $X$-band ($9.4$ GHz), and $Q$-band ($34$ GHz). Magnifying factors are listed on the left side of the corresponding spectrum lines. (c) ESR intensity ($\chi_{\mathrm{ESR}}$ ) and ESR $g$-values measured at $X$- and $Q$-band for $x=0.3$. $\chi_{\mathrm{ESR}}$ displays a Curie-like behavior (solid line) as expected for a local moment. The inset shows that at lower temperatures the evolution of  $\chi_{\mathrm{ESR}}$ is frequency/field dependent. The ESR $g$-values assume a constant value of $g=2$ (dashed line) for $T\geq60$ K.
\label{fig:spectraXQband}
}
\end{figure}

In  Fig. \ref{fig:spectraXQband}(a) we present the temperature ($T$) vs. Co-content ($x$) phase diagram for the series EuFe$_{2-x}$Co$_{x}$As$_{2}$, mostly based on our previous investigation \cite{nicklas_competition_2011}. 
The general properties of our ESR investigation for the EuFe$_{1.7}$Co$_{0.3}$As$_{2}$ sample are shown in Fig.~\ref{fig:spectraXQband}(b) (representative ESR spectra) and in Fig.~\ref{fig:spectraXQband}(c) [ESR intensity ($\chi_{\mathrm{ESR}}$), $g$-values].
The phase diagram shows that the SDW order, is fully suppressed for $x=0.2$. This is also the
only sample for which we could confirm a transition to a superconducting
phase [see $T_{\rm sc}$ symbol in Fig \ref{fig:spectraXQband}(a)]. Furthermore,
our magnetization measurements suggest that, for $x\geq0.2$, the Eu
local moment ordering changes from AFM to FM. These results are quite
similar to those obtained by the substitution of As by P 
\cite{jeevan_interplay_2011,ren_superconductivity_2009}, where superconductivity
is found only in a narrow region of the phase diagram and the nature
of the local order also changes on substitution. This similarity suggests that lattice
structural effects, rather than the effects of electronic doping,
play a fundamental role in tuning the phase diagram in these substitutional
studies \cite{rotter_different_2010}. However, more detailed investigations are required since
other works addressing the substitution of Fe by Co reported that
a transition to a superconducting state is realized in a much broader
interval of Co-substitution \cite{ying_electron_2010,chen_cobalt-doping_2010}. 

Direct inspection of the ESR spectra in Fig. \ref{fig:spectraXQband}(b) reveal that
the change in frequency corresponds to a change of the magnetic fields
applied to the sample. We point out that according to the resonance
condition $\hbar\omega=g\mu_{B}H_{\mathrm{res}}$ the typical fields,
for $g=2.0$, are roughly $340$ Oe for $L$-band, $3400$ Oe for
$X$-band and $12000$ Oe for $Q$-band. In addition, the $L$-band measurements are meaningful only in a relatively narrow temperature interval ($30\leq T\leq120$ K), since beyond this interval the broadening of the line makes the fitting process to be higly questionable. Furthermore, we will only discuss the $L$-band ESR linewidth ($\Delta H$).

$\chi_{\mathrm{ESR}}$ [Fig.\ref{fig:spectraXQband}(c)] is given by the the double integration of the observed spectra.
For $T\geq40$ K no differences between the $X$-band and $Q$-band
results can be resolved. In this temperature region, $\chi_{\mathrm{ESR}}$
follows a Curie-Weiss like behavior, indicated by the solid line in
the figure, as expected for well defined local moments. In the inset,
we show the low temperature data. The peak observed in the $X$-band
$\chi_{\mathrm{ESR}}$ marks the transition temperature of the local
moment magnetic order ($T_{\rm N}^{\mathrm{Eu}}$). This peak is not observed
in the $Q$-band $\chi_{\mathrm{ESR}}$ (i.e. at higher fields), suggesting
that the nature of the local moment magnetic ordering changes from
AFM to FM. However, it is indicating that some field needs to be applied
to fully stabilize the FM order in the region $x\geq0.2$ shown in
the phase diagram.

The $g$-values [Fig.~\ref{fig:spectraXQband}(c)] are obtained from the resonance condition. For $T\geq60$ K, they assume a constant value of $g=2$ that is close
to the Eu$^{2+}$ ionic value of $g=1.993$ found in insulators. At lower temperatures,
the $g$-values increase due to the onset of the internal fields associated
with the magnetic ordering of the local moments.

In metals, in the absence of a bottleneck effect, it is expected that the resonance position of a given paramagnetic ion is shifted with respect to its value in insulators \cite{barnes_theory_1981}. This $g$-shift ($\Delta g=g_{\mathrm{exp}}-g_{\mathrm{ins}}$) is given by

\begin{equation}
\Delta g=\left\langle \eta(E_{F})J(q=0)\right\rangle _{Av}=\eta(E_{F})\left\langle J(q=0)\right\rangle _{Av}\label{eq:deltag}\end{equation}
where in Eq. \ref{eq:deltag} $\left\langle J(q=0)\right\rangle _{Av}$
denotes an average over the Fermi surface of the $q=0$ component
of the exchange interaction $J(q)$, between the local moment and
the itinerant states, and $\eta(E_{F})$ is a constant DOS for a given spin direction at the Fermi surface (states eV$^{-1}$mol$^{-1}$spin$^{-1}$).
Therefore, one expects the $g$-values to reflect the evolution of
the Fermi surface and magnetic properties of the Fe$_{2-x}$Co$_{x}$As$_2$
layers, both supposed to be induced by Co-substitution. In contrast,
the $g$-values show nearly the same behavior for all measured samples,
indicating that they seem to reflect only the properties of the local
moment. This is a feature expected to be found in systems in a strong
bottleneck regime \cite{barnes_theory_1981,huber_electron_1976}.

Whereas the $g$-values usually reflect static properties, it is expected
that $\Delta H$ should be a direct probe of the dynamical behavior
of the electronic spin. Experiments on EuFe$_{2}$As$_{2}$ \cite{dengler_strong_2010,garcia_doping_2011}
single crystals have revealed two distinct behaviors of $\Delta H$. On one hand,
for $T>T_{\rm N}^{\mathrm{SDW}}$, a Korringa-like relaxation was found,
meaning that $\Delta H=\Delta H_{0}+b_{K}T$, where $\Delta H_{0}$
is the residual linewidth ($\Delta H$ at $T=0$) and $b_{K}$ is
the Korringa rate, which is determined by the exchange scattering
of the local moments (the 4$f$ Eu$^{2+}$ localized states) by the
itinerant electronic states (electronic states of FeAs layers). On
the other hand, for $T<T_{\rm N}^{\mathrm{SDW}}$, $\Delta H$ was shown
to display an angular dependence but was found to be nearly temperature
independent. This was discussed in terms of a suppressed Korringa relaxation
due to the gap opening at $T_{\rm N}^{\mathrm{SDW}}$.

These ideas were drawn on the basis that in metals, in the absence
of a bottleneck, the Korringa rate is given by \cite{barnes_theory_1981}:

\[
b_{K}=\frac{\pi k_{B}}{g\mu_{B}}\left\langle (N(E_{F})J(k_{F},k_{F}^{'}))^{2}\right\rangle _{Av}=\frac{\pi k_{B}}{g\mu_{B}}\eta(E_{F})^{2}\left\langle J(q)^{2}\right\rangle _{Av}\]

\begin{equation}
\equiv\frac{\pi k_{B}}{g\mu_{B}}\eta(E_{F})^{2}J^{2}\label{eq:bkorringa}\end{equation}
where $k_{B}$ is Boltzmann constant, $g$ is the measured $g$-value,
$\mu_{B}$ is the Bohr magneton, and $\left\langle J(q)^{2}\right\rangle _{Av}\equiv J^{2}$
is the average over the Fermi surface of the $q$-dependent exchange
coupling $J(q)$. In this picture, the gap opening at $T_{\rm N}^{\mathrm{SDW}}$
would make $\eta(E_{F})$ to vanish (or to decrease significantly),
thus changing the relaxation regime. The evolution of $\Delta H$
upon Co-substitution was also discussed in these terms, although details
were not given \cite{ying_electron_2010}.

The bottleneck effect does introduce an important modification in
Eq. \ref{eq:bkorringa}. In the bottleneck regime, the apparent Korringa
rate $b$ is given by \cite{barnes_theory_1981}:
\begin{equation}
b=(\frac{1/T_{\mathrm{ceL}}}{1/T_{\mathrm{ceI}}+1/T_{\mathrm{ceL}}})b_{K}\label{eq:bbottleneck}\end{equation}
In the strong bottleneck regime, the exchange scattering will no longer
determine the relaxation (at least not in first order), since $1/T_{\mathrm{ceI}}\propto J^{2}$.
The expression for $\Delta H$ is usually casted in the following
form \cite{barnes_theory_1981}:
\begin{equation}
\Delta H=\Delta H_{0}+\frac{\chi_{\mathrm{e}}}{\chi_{\mathrm{dc}}}\frac{1}{T_{\mathrm{ceL}}}\label{eq:deltaHbottle}\end{equation}
where $\chi_{\mathrm{e}}$ is the susceptibility of the itinerant
electronic system (usually the Pauli susceptibility), and $\chi_{\mathrm{dc}}$
is the measured $dc$-susceptibility of the local moment. It is noteworthy
that Eq. \ref{eq:deltaHbottle} mimics the Korringa relaxation but,
in contrast, depends strongly on the concentration of the ESR probe.
A similar expression can also be deduced by the general arguments given by Huber \cite{huber_electron_1976}.
In the following, we shall explain how Eq. \ref{eq:bbottleneck} and
in particular Eq. \ref{eq:deltaHbottle} are more appropriate than
Eq. \ref{eq:bkorringa} to address the existing experimental data. 

\begin{figure}
\begin{centering}
\includegraphics[scale=0.4]{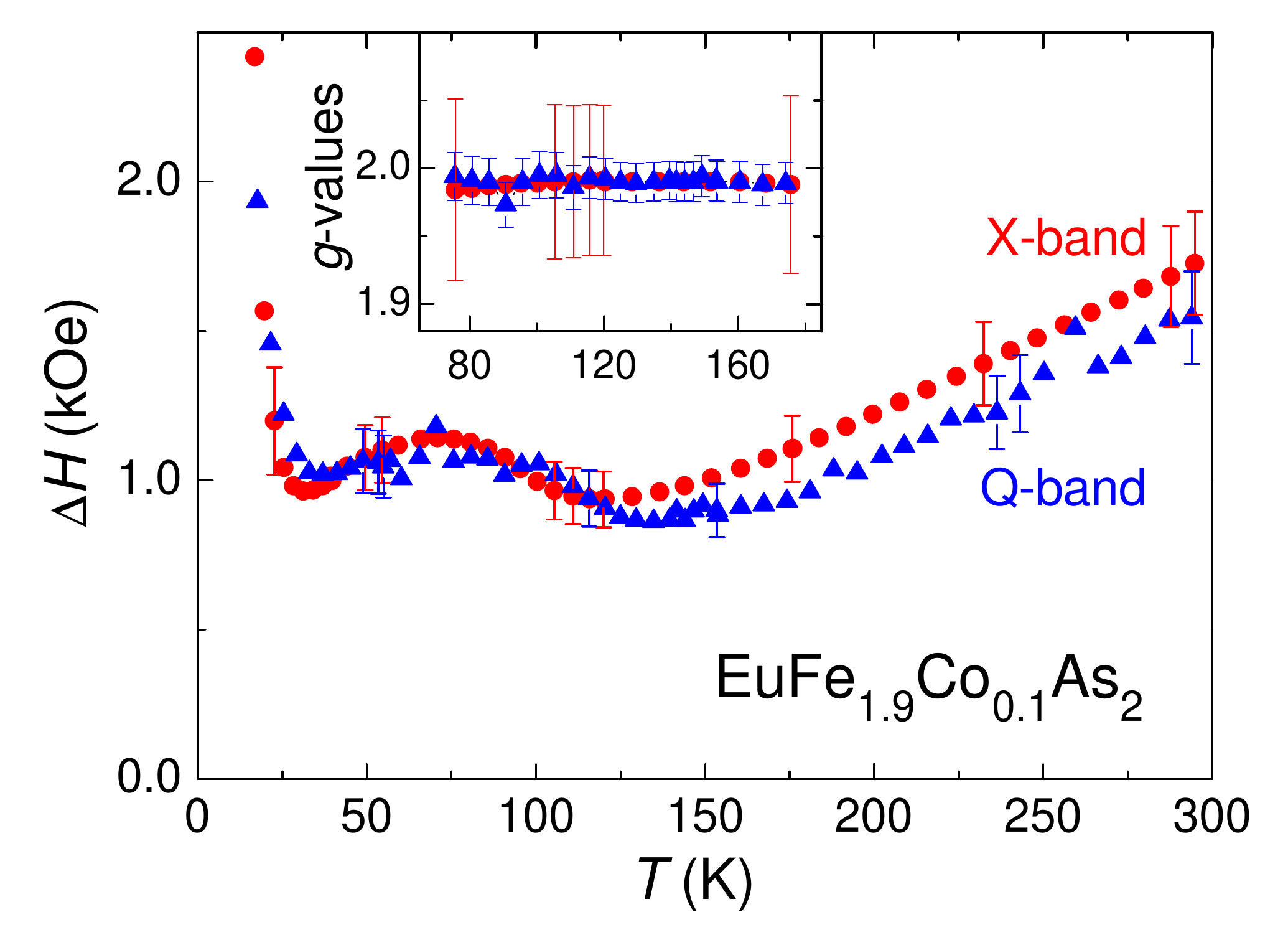}
\par\end{centering}

\caption{(Color online) The evolution of the  linewidth ($\Delta H$) as a
function of $T$ obtained for EuFe$_{1.9}$Co$_{0.1}$As$_{2}$, which
undergoes a SDW transition at $T_{\rm N}^{SDW}=100$~K. At this temperature, in
contrast to all other samples, $\Delta H$ no longer follows a ``Korringa-like''
behavior (see text). The inset shows in detail that the $g$-values,
unlike $\Delta H$, are unaffected by this transition at $100$ K.\label{fig:x01lwQXgva}}

\end{figure}

We begin discussing Fig.~\ref{fig:x01lwQXgva}, which shows the $X$-band
and $Q$-band linewidth and $g$-values (in the inset) for the EuFe$_{1.9}$Co$_{0.1}$As$_{2}$
sample, which is the only sample investigated diplaying a SDW transition. 
We confirm a linear increase of $\Delta H$ for $T>T_{\rm N}^{SDW}=100$
K, and a dome-like behavior of $\Delta H$ \cite{garcia_doping_2011}
between $40\leq T\leq100$ K (not present in Ref. \cite{dengler_strong_2010}).
The $g$-values (see inset) are insensitive to the change in the relaxation
regime. Furthermore, the relaxation is slightly slower at $Q$-band.

In principle, both expressions (Eqs. \ref{eq:bbottleneck} and \ref{eq:deltaHbottle})
may be used to explain the behavior of $\Delta H$. For $T>T_{\rm N}^{\mathrm{SDW}}$,
one could consider that the linear increase of $\Delta H$ originates
from a reduced Korringa process or from the inverse of the high temperature
$\chi_{dc}$. For $T<T_{\rm N}^{\mathrm{SDW}}$, the dome-like behavior
of $\Delta H$ finds no explanation in the Korringa picture. It could
be that it is a coherence peak, due to the gap opening of the SDW
transition or, due its sample dependence, it could be ascribed to
the dome which is observed in the resistivity \cite{nicklas_competition_2011}.
The resistivity is reflected through $1/T_{\mathrm{ceL}}$, that comprises
all of the scattering processes (energy dissipation) of the itinerant
system. The insensitivity of the $g$-values to the changes of the
relaxation regime indicates a strong relaxation bottleneck. 

\begin{figure}
\begin{centering}
\includegraphics[scale=0.4]{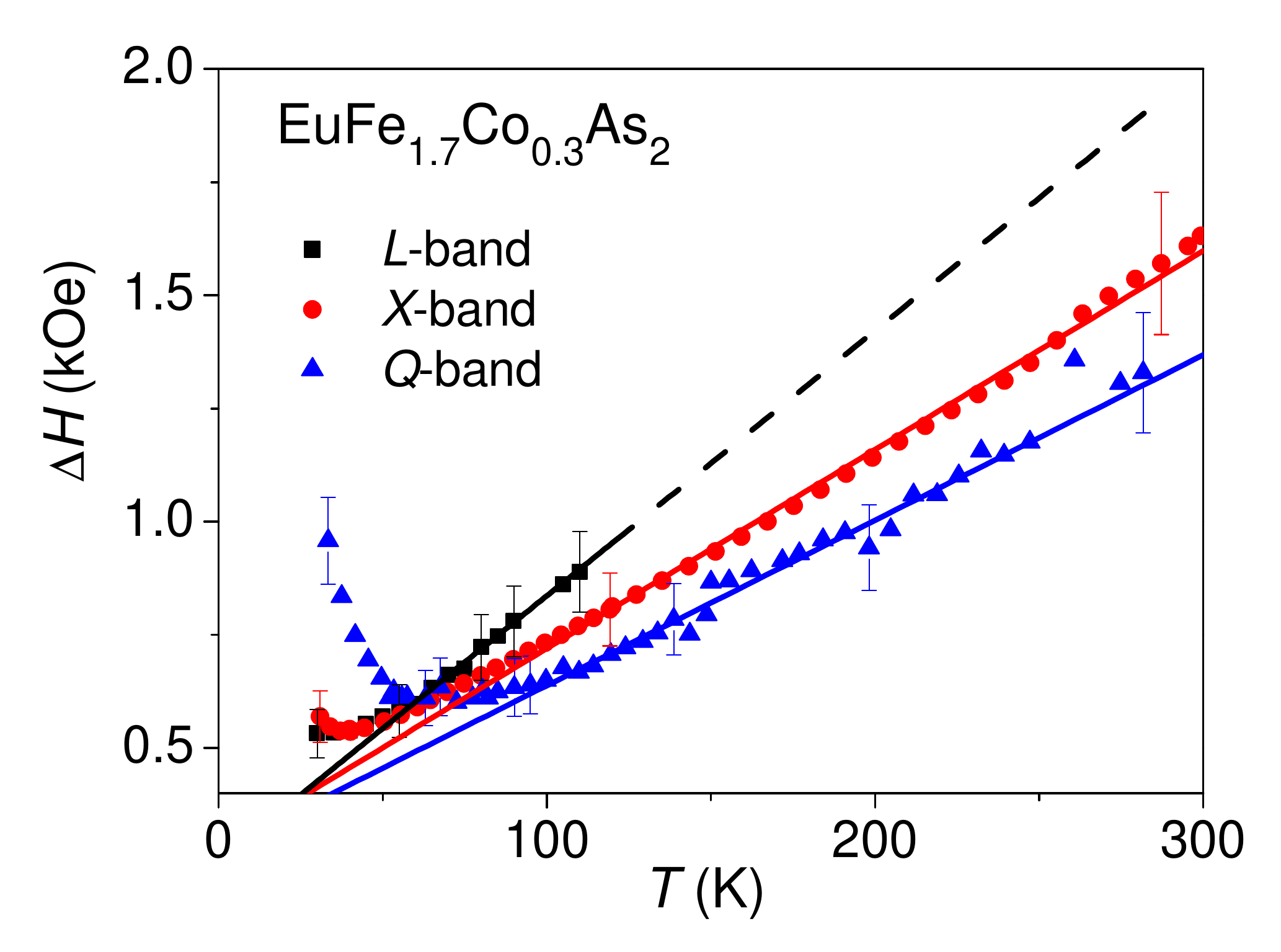}
\par\end{centering}

\caption{(Color online) Temperature dependence of the linewidth ($\Delta H$)
of EuFe$_{1.7}$Co$_{0.3}$As$_{2}$ for $L$-band, $X$-band, and
$Q$-band frequencies. The data are representative for all measured,
but the $x=0.1$, samples (for $x=0.1$ see Fig. \ref{fig:x01lwQXgva}).
The solid lines are the best linear fit to the data taken at $X$-band
and $Q$-band above $T=100$ K. The dashed line is an extrapolation
of the low temperature fit of $L$-band data. \label{fig:x03XQ075lw}}

\end{figure}

For all other measured samples (with $x>0.1$), the evolution of $\Delta H$,
as a function of temperature and frequency/field, is well represented
by the data in Fig.~\ref{fig:x03XQ075lw}, which displays the $L$-band,
$X$-band, and $Q$-band linewidth for EuFe$_{1.7}$Co$_{0.3}$As$_{2}$.
For $T\geq40$ K $\Delta H$ broadens linearly with increasing temperature
whereas in the low temperature region, with decreasing temperature,
the proximity of the magnetic phase transition leads to the observed
fast broadening of $\Delta H$. The linear increase was fitted to
the expression $\Delta H=\Delta H_{0}+bT$ for $80\,{\rm K}\leq T\leq120$
K in the case of the $L$-band data and for  $100\,{\rm K} \leq T\leq300$ K in the
cases of the  $X$- and $Q$-band data. The dashed line shows
an extrapolation for $T\geq120$ K of the $L$-band fitting result.

\begin{figure}
\begin{centering}
\includegraphics[scale=0.30]{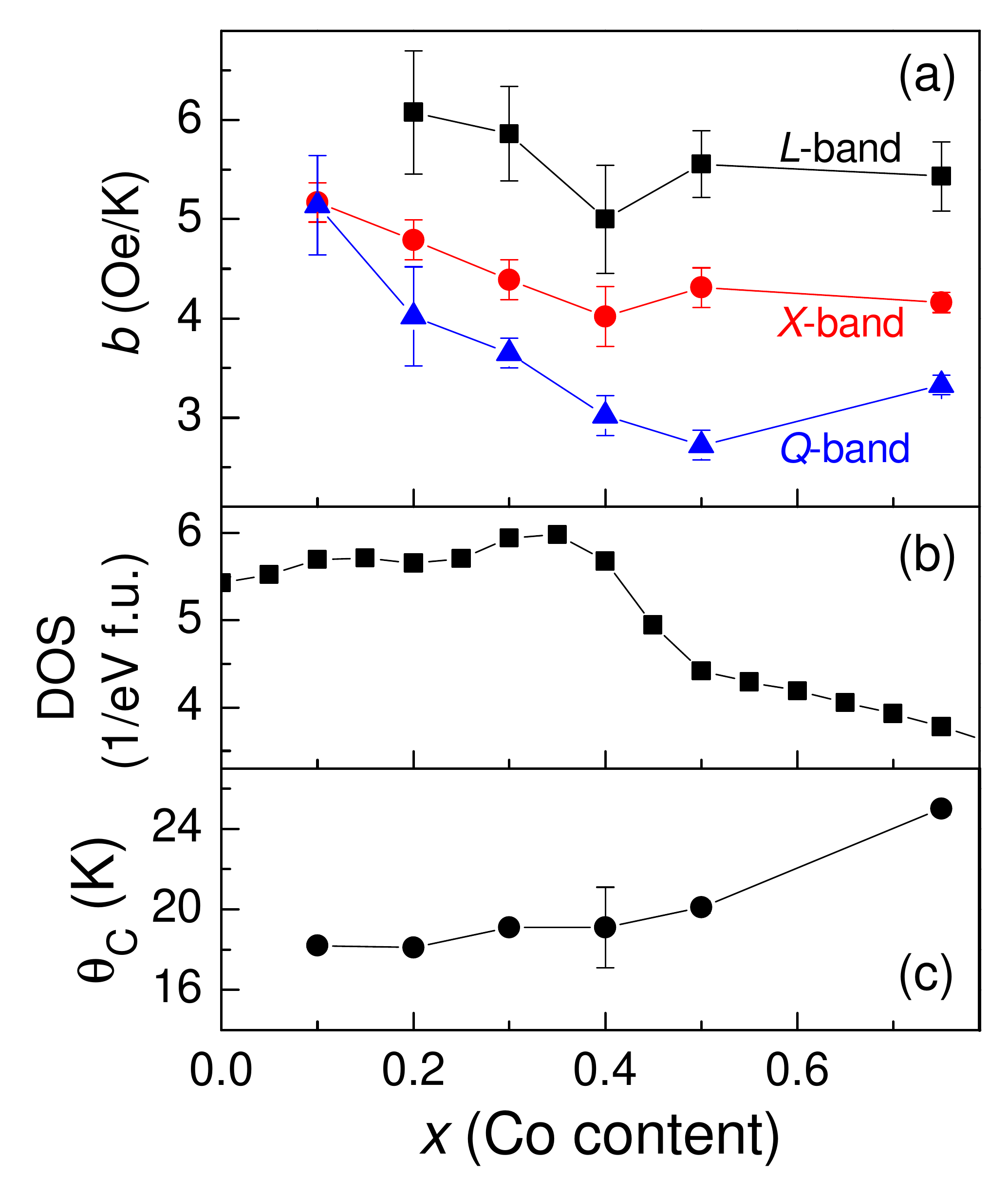} 
\par\end{centering}

\caption{(Color online) Evolution as a function of Co-substitution of (a) the
$b$ parameter, obtained at different frequencies/fields, extracted
from a linear fitting of $\Delta H$ as a function of temperature,
(b) the calculated total density of states (DOS), and (c) the Curie temperature $\theta_{C}$ obtained from
dc-susceptibility measurements. \label{fig:bqtdsdopingevol}}

\end{figure}

The obtained $b$ parameters are compiled in Fig. \ref{fig:bqtdsdopingevol}(a)
as a function of frequency/field and Co-substitution. As a function
of $x$, the $b$ parameter decreases along the series and is nearly
constant for $x>0.4$. With increasing the applied frequency/field
the $b$ parameter is clearly reduced. This is best pronounced in
the region $x\ge0.2$ where instead of the now fully suppressed ordered
phase of the itinerant subsystem, one expects strong magnetic fluctuations.

The observed behavior of the $b$ parameter on the frequency, being
understood as an effect of the frequency-equivalent magnetic field,
can be explained by the dependence of the $b$ parameter on $1/T_{\mathrm{ceL}}$.
In a magnetic compound, the itinerant subsystem will also dissipate
energy through a magnetic process, meaning that $1/T_{\mathrm{ceL}}$
is also sensitive to spin-spin scattering. A relatively higher field
suppresses the magnetic fluctuations in the system thus lowering the
scattering rate of the conduction electrons by the magnetic fluctuations
(spin-spin scattering). Therefore, as observed [Fig. \ref{fig:bqtdsdopingevol}(a)],
at higher fields $b$ should be smaller.

The evolution of the $b$ parameter as a function of the Co-substitution
is a somewhat more evolved subject and a number of distinct effects
should be considered. First, we consider that the the Co-substitution contributes to the DOS at the Fermi
level [$\eta(E_{F})$] by electronic doping. In the case of a relaxation
dominated by the Korringa process, $b$ should be proportional to
$\eta(E_{F})^{2}$ (see Eq.~\ref{eq:bkorringa}). To further investigate
this point, we performed band structure calculations to reveal the
change of $\eta(E_{F})$ upon Co-substitution. It has been shown previously
that the effect of Co-substitution on the SDW, which is closely related
to the Fermi surface, can be well described using the VCA \cite{kasinathan_afe2as2_2009}. The result for $0<x<0.8$
is presented in Fig.~\ref{fig:bqtdsdopingevol}(b). $\eta(E_{F})$
shows a small upwards variation for $x$ up to $\approx0.4$ , and
then starts to decrease continuously for higher values of $x$.
This behavior is nearly opposite to the one observed in $b$. 

The evolution of the local moment fluctuations, as expressed by the Curie temperature $\theta_{C}$, along the series is presented in Fig. \ref{fig:bqtdsdopingevol}(c). 
Due to the effects of exchange narrowing, $\Delta H$ would be narrower for larger fluctuations. Accordingly, $b$ should decrease monotonically as $\theta_{C}$ increases. However, this is not the observed behavior, suggesting that the local moment fluctuations alone cannot
explain the evolution of the Eu$^{2+}$ spin relaxation as function of Co-substitution.

\begin{figure}
\begin{centering}
\includegraphics[scale=0.27]{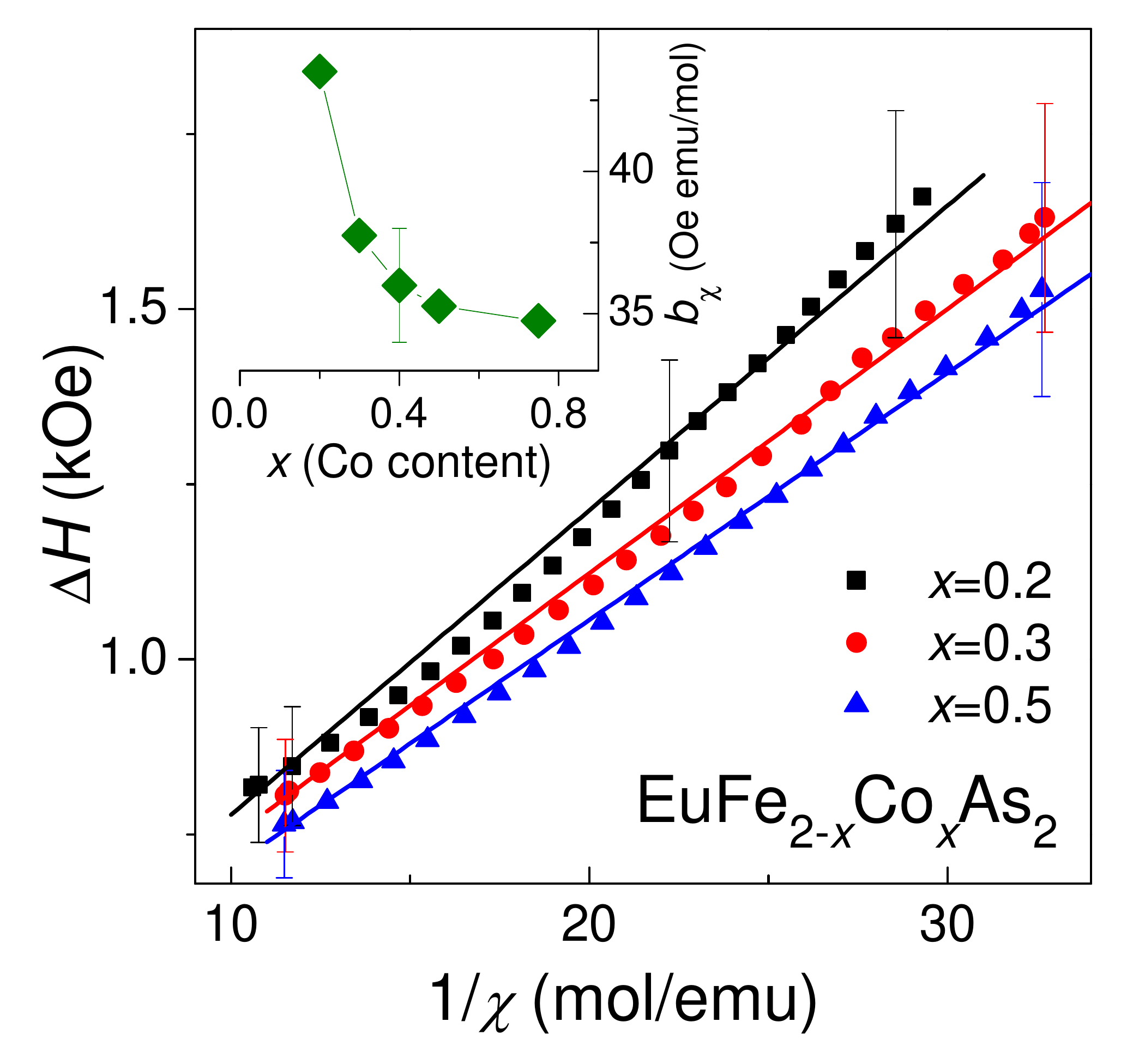}
\par\end{centering}

\caption{(Color online)  Plot of $\Delta H$ as a function of reciprocal susceptibility $1/\chi_{dc}$
($H=1$ kOe,  $T\geq100$ K). The solid line is the linear fitting of the data
and the obtained coefficient $b_{\chi}$ is presented as a function
of Co-substitution in the inset.\label{fig:figsuscdeltaH}}

\end{figure}

In Fig.~\ref{fig:figsuscdeltaH} we present some relevant information
to this discussion. We follow the suggestion of Eq. \ref{eq:deltaHbottle}
and show a plot of $\Delta H$ as function of $1/\chi_{dc}$. It is
clearly seen that for a broad temperature interval, $T\geq100$ K,
$\Delta H$ is proportional to $1/\chi_{dc}$. The slopes of these
curves ($b_{\chi}$), for the whole set of samples (but the $x=0.1$
sample), are compiled in the inset of the figure. A good qualitative 
correlation between the concentration dependence of the $b$ parameter 
and the results of the slopes $b_{\chi}$ is found (see inset of Fig.~\ref{fig:figsuscdeltaH}). Following Eq. \ref{eq:deltaHbottle}, 
this slope should be written as $b_{\chi}\approx\chi_{e}(1/T_{ceL})$, 
meaning that it expresses an interplay between the magnetism of
the itinerant states in the Fe$_{2-x}$Co$_{x}$As$_2$ layers ($\chi_{e}$)
and the multiple scattering processes taking place in the system ($1/T_{ceL}$). 

On a technical level, the above results when put together strongly suggest that the Korringa process has a marginal role in the spin relaxation in these systems. This means that the Eu$^{2+}$ spin relaxation is driven by the indirect RKKY coupling among the Eu$^{2+}$ spins, that is under the influence of the magnetic fluctuations of the itinerant subsystem.

The field dependence of the relaxation reveals the presence of strong magnetic
fluctuations after the total suppression of the AFM order of the itinerant
subsystem. For $x>0.4$, where $b$ is nearly constant as a function
of $x$, the field dependence of $b$ is strongly pronounced, suggesting
stronger magnetic fluctuations than for $x<0.4$. Furthermore, the field 
dependence or the relaxation testifies to the existence of magnetic 
fluctuations even at relatively high levels of Co-substitution in
the Fe$_{2-x}$Co$_{x}$As$_2$ layers.

As already noted, the suppression of the superconducting state by
the pair-breaking effect of the Eu$^{2+}$ moments in optimally doped
Sr$_{1-y}$Eu$_{y}$Fe$_{2-x}$Co$_{x}$As$_{2}$ was recently addressed
\cite{hu_phase_2011} and here we use the data of Ref.~\cite{hu_phase_2011} in the subsequent
discussion. Pair breaking by an paramagnetic impurity in a superconductor
is described by the Abrikosov-Gorkov expression and is driven by an
effective exchange scattering that was demonstrated to be comparable
with those determined by ESR \cite{davidov_electron_1973} (see
Eq. \ref{eq:bkorringa}). The required DOS for this
calculation was shown to be nearly the same for SrFe$_{2}$As$_{2}$
and EuFe$_{2}$As$_{2}$, and amounts to $\eta(E_{F})=0.8$ states/eV.f.u.
\cite{jeevan_electrical_2008}. The estimated exchange scattering
is $J=6.1$ meV which, by Eq. \ref{eq:bkorringa}, would drive a relaxation
no faster than $1$ Oe/K. This should be taken as an upper bound for
the value of $b_{K}$ if the relaxation were determined by the Korringa
process. Therefore, this result provides further evidence for the
above described scenario where the relaxation is driven by the exchange
coupling among the local spins \cite{huber_electron_1976}. In the
case of a more traditional bottleneck effect \cite{rettori_electron-spin_1973,rettori_dynamic_1974,barnes_theory_1981},
the magnetic fluctuations could increase the apparent relaxation rate
by partially, or completely, opening the bottleneck. However, the
Korringa rate $b_{K}$ could not exceed the value of $b_{K}=1$ Oe/K
estimated for a full Korringa process. Although not conclusive, this
is another important point in favor of an exchange coupled spins scenario
to interpret the bottleneck in EuFe$_{2-x}$Co$_{x}$As$_{2}$.

\section{Conclusions}

From investigating the field dependence of the Eu$^{2+}$ spin resonance in the series EuFe$_{2-x}$Co$_{x}$As$_{2}$ we found that the electronic properties of the itinerant subsystem are directly reflected in the ESR relaxation mechanism. We have shown that this is achieved in terms of a so-called relaxation bottleneck of exchange coupled Eu$^{2+}$ spins. In particular, this kind of relaxation mechanism is suggested by the opposing evolutions of the linewidth slope ($b$) and the calculated DOS at the Fermi level along the series. 

The clear field dependence of the relaxation indicates the presence of magnetic fluctuations and provides a picture of how the relative strength of the itinerant magnetic fluctuations evolves along the series. Interestingly, this field dependence is well pronounced for $x\geq0.2$ where the AFM order of the itinerant subsystem is suppressed, indicating it is a property of the emerging itinerant magnetic fluctuations. Furthermore, the field effect is visible even at $x = 0.75$ which is far away from  $x = 0.2$.

The ESR results add further evidence that the presence of the Eu$^{2+}$ spins nearly does not affect the electronic or magnetic properties of the itinerant subsystem. Therefore, one could expect that our picture on the evolution of the magnetic fluctuations is a general property of the Fe$_{2-x}$Co$_{x}$As$_2$ layers.

\section*{Acknowledgments}

We are in great debt to Deepa Kasinathan for many discussions on the
properties of the EuFe$_{2}$As$_{2}$ system. Part of this work has
been performed within the framework of DFG SPP$1458$.

\section*{References}

\begin{thebibliography}{10}
\expandafter\ifx\csname url\endcsname\relax
  \def\url#1{{\tt #1}}\fi
\expandafter\ifx\csname urlprefix\endcsname\relax\def\urlprefix{URL }\fi
\providecommand{\eprint}[2][]{\url{#2}}

\bibitem{kamihara_iron-based_2008}
Kamihara Y, Watanabe T, Hirano M and Hosono H 2008 {\em J. Am. Chem. Soc.\/}
  {\bf 130} 3296--3297 ISSN 0002-7863

\bibitem{zhi-an_superconductivity_2008}
{Zhi-An} R, Wei L, Jie Y, Wei Y, {Xiao-Li} S, {Zheng-Cai}, {Guang-Can} C,
  {Xiao-Li} D, {Li-Ling} S, Fang Z and {Zhong-Xian} Z 2008 {\em Chinese Physics
  Letters\/} {\bf 25} 2215--2216 ISSN {0256-307X}, 1741-3540

\bibitem{johnston_puzzle_2010}
Johnston D~C 2010 {\em Advances in Physics\/} {\bf 59} 803--1061 ISSN 00018732

\bibitem{monthoux_superconductivity_2007}
Monthoux P, Pines D and Lonzarich G~G 2007 {\em Nature\/} {\bf 450} 1177--1183
  ISSN 0028-0836

\bibitem{saxena_superconductivity_2000}
Saxena S~S, Agarwal P, Ahilan K, Grosche F~M, Haselwimmer R~K~W, Steiner M~J,
  Pugh E, Walker I~R, Julian S~R, Monthoux P, Lonzarich G~G, Huxley A, Sheikin
  I, Braithwaite D and Flouquet J 2000 {\em Nature\/} {\bf 406} 587--592 ISSN
  0028-0836

\bibitem{le_tacon_intense_2011}
Le~Tacon M, Ghiringhelli G, Chaloupka J, Sala M~M, Hinkov V, Haverkort M~W,
  Minola M, Bakr M, Zhou K~J, {Blanco-Canosa} S, Monney C, Song Y~T, Sun G~L,
  Lin C~T, De~Luca G~M, Salluzzo M, Khaliullin G, Schmitt T, Braicovich L and
  Keimer B 2011 {\em Nat Phys\/} {\bf 7} 725--730 ISSN 1745-2473

\bibitem{jeevan_electrical_2008}
Jeevan H, Hossain Z, Kasinathan D, Rosner H, Geibel C and Gegenwart P 2008 {\em
  Phys. Rev. B\/} {\bf 78} ISSN 1098-0121, {1550-235X}

\bibitem{ren_antiferromagnetic_2008}
Ren Z, Zhu Z, Jiang S, Xu X, Tao Q, Wang C, Feng C, Cao G and Xu Z 2008 {\em
  Phys. Rev. B\/} {\bf 78} 052501

\bibitem{kasinathan_afe2as2_2009}
Kasinathan D, Ormeci A, Koch K, Burkhardt U, Schnelle W and Rosner H 2009 {\em
  New journal of physics\/} {\bf 11} ISSN 1367-2630

\bibitem{xiao_magnetic_2009}
Xiao Y, Su Y, Meven M, Mittal R, Kumar C~M~N, Chatterji T, Price S, Persson J,
  Kumar N, Dhar S~K, Thamizhavel A and Brueckel T 2009 {\em Phys. Rev. B\/}
  {\bf 80} 174424

\bibitem{jeevan_high-temperature_2008}
Jeevan H~S, Hossain Z, Kasinathan D, Rosner H, Geibel C and Gegenwart P 2008
  {\em Phys. Rev. B\/} {\bf 78} 092406

\bibitem{ying_electron_2010}
Ying J~J, Wu T, Zheng Q~J, He Y, Wu G, Li Q~J, Yan Y~J, Xie Y~L, Liu R~H, Wang
  X~F and Chen X~H 2010 {\em Phys. Rev. B\/} {\bf 81} 052503

\bibitem{jeevan_interplay_2011}
Jeevan H, Kasinathan D, Rosner H and Gegenwart P 2011 {\em Phys. Rev. B\/} {\bf
  83} ISSN 1098-0121, {1550-235X}

\bibitem{ren_superconductivity_2009}
Ren Z, Tao Q, Jiang S, Feng C, Wang C, Dai J, Cao G and Xu Z 2009 {\em Phys.
  Rev. Lett.\/} {\bf 102} ISSN 0031-9007

\bibitem{miclea_evidence_2009}
Miclea C~F, Nicklas M, Jeevan H~S, Kasinathan D, Hossain Z, Rosner H, Gegenwart
  P, Geibel C and Steglich F 2009 {\em Phys. Rev. B\/} {\bf 79} 212509

\bibitem{kurita_phase_2011}
Kurita N, Kimata M, Kodama K, Harada A, Tomita M, Suzuki H~S, Matsumoto T,
  Murata K, Uji S and Terashima T 2011 {\em Phys. Rev. B\/} {\bf 83} 214513

\bibitem{hu_phase_2011}
Hu R, Budko S, Straszheim W and Canfield P 2011 {\em Phys. Rev. B\/} {\bf 83}
  094520

\bibitem{zhou_high-resolution_2010}
Zhou B, Zhang Y, Yang L, Xu M, He C, Chen F, Zhao J, Ou H, Wei J, Xie B, Wu T,
  Wu G, Arita M, Shimada K, Namatame H, Taniguchi M, Chen X~H and Feng D~L 2010
  {\em Phys. Rev. B\/} {\bf 81} 155124

\bibitem{terashima_magnetotransport_2010}
Terashima T, Kurita N, Kikkawa A, Suzuki H~S, Matsumoto T, Murata K and Uji S
  2010 {\em J. Phys. Soc. Jpn.\/} {\bf 79} 103706 ISSN 0031-9015

\bibitem{nicklas_competition_2011}
Nicklas M, Kumar M, Lengyel E, Schnelle W and {Leithe-Jasper} A 2011 {\em
  Journal of Physics: Conference Series\/} vol 273 p 012101

\bibitem{dengler_strong_2010}
Dengler E, Deisenhofer J, Krug~von Nidda H, Khim S, Kim J~S, Kim K~H, Casper F,
  Felser C and Loidl A 2010 {\em Phys. Rev. B\/} {\bf 81} ISSN 1098-0121

\bibitem{barnes_theory_1981}
Barnes S 1981 {\em Advances in Physics\/} {\bf 30} 801--938 ISSN 0001-8732

\bibitem{van_vleck_dipolar_1948}
Van~Vleck J~H 1948 {\em Phys. Rev.\/} {\bf 74} 1168--1183

\bibitem{huber_electron_1976}
Huber D~L 1976 {\em Phys. Rev. B\/} {\bf 13} 291--294

\bibitem{pascher_magnetic_2010}
Pascher N, Deisenhofer J, von Nidda H~A, Hemmida M, Jeevan H~S, Gegenwart P and
  Loidl A 2010 {\em Arxiv preprint {arXiv:1001.1302}\/}

\bibitem{garcia_doping_2011}
Garcia F~A, Bittar E~M, Adriano C, Garitezi T~M, Rettori C and Pagliuso P~G
  2011 {\em Journal of Physics: Conference Series\/} {\bf 273} 012093 ISSN
  1742-6596

\bibitem{rettori_dynamic_1974}
Rettori C, Kim H~M, Chock E~P and Davidov D 1974 {\em Phys. Rev. B\/} {\bf 10}
  1826

\bibitem{rettori_electron-spin_1973}
Rettori C, Davidov D, Orbach R, Chock E~P and Ricks B 1973 {\em Phys. Rev. B\/}
  {\bf 7} 1

\bibitem{joshi_analysis_2004}
Joshi J~P and Bhat S~V 2004 {\em J. Magn. Reson.\/} {\bf 168} 284--287 ISSN
  1090-7807

\bibitem{wykhoff_local_2007}
Wykhoff J, Sichelschmidt J, Lapertot G, Knebel G, Flouquet J, Fazlishanov I~I,
  Krug~von Nidda H, Krellner C, Geibel C and Steglich F 2007 {\em Science and
  Technology of Advanced Materials\/} {\bf 8} 389--392 ISSN 1468-6996

\bibitem{opahle_full-potential_1999}
Opahle I, Koepernik K and Eschrig H 1999 {\em Phys. Rev. B\/} {\bf 60}
  14035--14041

\bibitem{perdew_accurate_1992}
Perdew J~P and Wang Y 1992 {\em Phys. Rev. B\/} {\bf 45} 13244--13249

\bibitem{kasinathan_electronic_2012}
Kasinathan D, Wagner M, Koepernik K, {Cardoso-Gil} R, Grin Y and Rosner H 2012
  {\em Phys. Rev. B\/} {\bf 85} 035207

\bibitem{uhoya_anomalous_2010}
Uhoya W, Tsoi G, Vohra Y~K, {McGuire} M~A, Sefat A~S, Sales B~C, Mandrus D and
  Weir S~T 2010 {\em J. Phys.: Condens. Matter\/} {\bf 22} 292202 ISSN
  0953-8984, {1361-648X}

\bibitem{rotter_different_2010}
Rotter M, Hieke C and Johrendt D 2010 {\em Phys. Rev. B\/} {\bf 82} 014513

\bibitem{chen_cobalt-doping_2010}
Chen X, Ren Z, Ding H and Liu L 2010 {\em {SCIENCE} {CHINA} Physics, Mechanics
  \& Astronomy\/} {\bf 53} 1212--1215

\bibitem{davidov_electron_1973}
Davidov D, Chelkowski A, Rettori C, Orbach R and Maple M~B 1973 {\em Phys. Rev.
  B\/} {\bf 7} 1029

\end{thebibliography}

\providecommand{\newblock}{}

\end{document}